\documentstyle[twoside,fleqn,espcrc2,epsf]{article}
\input psfig

\title{Improving the Lattice QED Action}

\author{Simon Hands\address{Department of
Physics, University of Wales, Swansea, \\
        Singleton Park, Swansea, SA2 8PP, U.K.}
\hfill SWAT/94/54\\
Talk presented at {\sl Lattice '94\/}, Bielefeld, Germany.\hfill
hep-lat/9411049
}

\begin{document}

\begin{abstract}
Strongly coupled QED is a model whose physics is dominated by
short-ranged effects. In order to assess which features of numerical
simulations of the chiral phase transition are universal and which are
not, we have formulated a quenched version of the model in which photon
degrees of freedom are defined on a lattice of spacing $a$, but fermions
only on a lattice of spacing $2a$. The fermi-photon interaction is then
obtained via a blocking procedure, whose parameters allow a degree of
control over the relative importance of short wavelength modes. Results
from a variety of models are presented; the critical exponents $\delta$
and $\beta_m$ governing the transition appear to be independent  of the
blocking, or even of whether a gauge-invariant action is used for the
photons.
\end{abstract}

\maketitle

In this talk I will describe work done in collaboration with John Kogut.
We have been experimenting with different variants of the non-compact
lattice QED action. The motivation is twofold: firstly we wish to get a
better understanding of chiral symmetry breaking, which occurs at strong
coupling, and in particular to disentangle the role of various features,
eg. whether the physics is dominated by short or long ranged effects;
the relative importance of photon exchange and four-fermi contact
interactions; and the role, if any, of magnetic monopole excitations.
Secondly, if there were a UV fixed point in strongly coupled QED, then
there should be the possibility of some renormalization group
``improvement'' which would widen the scaling region seen in simulations
and give cleaner signals.

A specific example is the hypothesis that the fixed point can be
described by the gauged Nambu -- Jona-Lasinio model:
\begin{equation}
{\cal L}=\bar\psi D{\!\!\!\! /}\,\psi+{1\over4}F_{\mu\nu}^2
+G[(\bar\psi\psi)^2-(\bar\psi\gamma_5\psi)^2].
\end{equation}
Studies using the quenched ladder approximation \cite{HKK}\cite{MBL}
predict a line of fixed
points running from the NJL mean field transition at $(\alpha,G)=(0,4)$
to the essential singularity discussed by Miransky and Bardeen {\it et
al\/} at $(\alpha_c,1)$ with $\alpha_c=\pi/3$. Critical exponents of the
chiral transition are analytic functions of $\alpha/\alpha_c$ along the
fixed line. Numerical simulations of quenched non-compact QED (NC QED) yield
estimates of exponents consistent with crossing the fixed line at
$(0.44\alpha_c,3.06)$ \cite{HKK}.
Why might the gauged NJL model be a good
effective theory for lattice QED? One clue is the lattice photon
propagator. In Feynman gauge,
\begin{equation}
D_{\mu\nu}^{-1}(k)=\delta_{\mu\nu}\sum_\rho4\sin^2{k_\rho\over2}.
\end{equation}
This falls below the continuum form $k^2$ at the edge of the Brillouin
zone, and generates an effective contact interaction between vector
currents, as well as the term due to photon exchange.

Our proposal is to modify the relative weights of large and small $k$
modes in the propagator by modifying the lattice photon action. The
approach is simple: from photon fields $\theta_\mu(x)$ generated on a
$(2N)^4$ lattice using the standard non-compact action we form blocked
links $\Theta_\mu(x)$ on a $N^4$ lattice via:

\vbox{
\psfig{figure=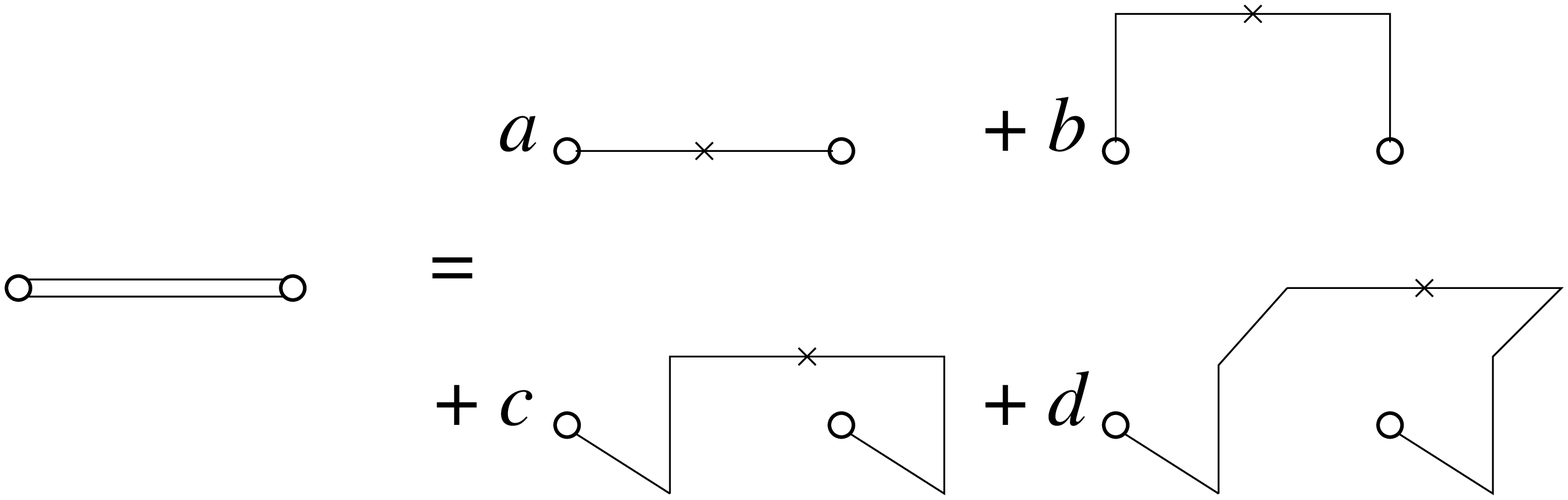,width=2.8in}
}
The action remains Gaussian under blocking and it is straightforward to
examine the new dispersion curves (fig. 1). From the large family of
possible blocked actions we chose two forms to examine in simulations:
\begin{figure}
\vbox{\psfig{figure=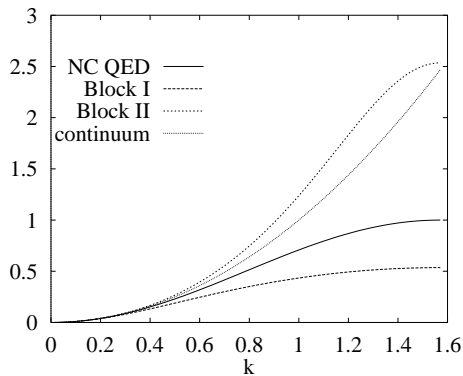,width=2.8in}}
\caption{Inverse photon propagator parallel to (0,0,0,1)}
\end{figure}
(I) $a=1,b=c=d=0$. We found this to enhance short wavelength modes over
long, effectively increasing the ``screening'' in the model; and (II)
$a=1,b={1\over8},c=d={1\over16}$. This choice gave a relative
suppression of short wavelength modes. For comparison we also tried a
``totally screened'' random link (RL)
model in which the $\{\Theta\}$ are generated
independently on each link with Gaussian weight
$\exp(-16\Theta_\mu^2)$.

To examine the critical properties of the chiral transition, we
estimated the chiral condensate $\langle\bar\psi\psi\rangle$ in the
chiral limit bare mass $m\to0$ on (blocked)
lattice sizes $8^4$, $12^4$, and $16^4$ in two complementary ways \cite{HKK}.
First we measure $\langle\bar\psi\psi\rangle$ as a function of
$m$ by inverting $(D{\!\!\!\! /}\,[\Theta]+m)$, where $D{\!\!\!\! /}\,$
is the staggered fermion covariant derivative, using a conjugate
gradient alogorithm. We used a sequence of $m$ values
$0.005,0.004,\ldots,0.001$. Secondly, we estimate the spectral density
function $\rho(\lambda)$ by using the Lanczos algorithm to find the
eigenvalues of $D{\!\!\!\! /}\,[\Theta]$. The spectral density is
related to a branch cut in $\langle\bar\psi\psi(m)\rangle$ along the
imaginary $m$ axis:
\begin{equation}
2\pi\rho(\lambda)=
\displaystyle\lim_{\varepsilon\to0}
\langle\bar\psi\psi(-i\lambda+\varepsilon)\rangle\!-\!
\langle\bar\psi\psi(-i\lambda-\varepsilon)\rangle.
\end{equation}
On the assumption  of power-law scaling at criticality, we have a
relation between the critical coupling $\beta_c\equiv1/e_c^2$ and the
critical exponent $\delta$:
\begin{equation}
\langle\bar\psi\psi(m)\rangle\vert_{\beta=\beta_c}=Am^{1/\delta}.
\end{equation}
The two parameters can be estimated by looking for the best fit of eqn.
(4) as a function of $\beta$ to the conjugate gradient data.
Alternatively, the Lanczos data can be fitted using the ansatz
\begin{equation}
\rho(\lambda)\vert_{\beta=\beta_c}=B\lambda^{1/\delta'}.
\end{equation}
As a cross check, using (3) we have the relations
\begin{equation}
\delta=\delta'\;;\;A={{\pi B}\over{\cos(\pi/2\delta)}}.
\end{equation}
By demanding consistency between the two sets of measurements, we can
achieve accuracy of better than a percent in estimating $\beta_c$, and
5\% in $\delta$. It is also possible to estimate the magnetic exponent
$\beta_m$ from the $\rho(\lambda)$ data, though with less precision:
\begin{equation}
\rho(\lambda=0)=C(\beta_c-\beta)^{\beta_m}.
\end{equation}

Another observable we kept track of was the size of the largest
connected monopole cluster, along with the associated cluster
susceptibility \cite{HW}. By monitoring this on a series of lattice sizes
$8^4,10^4,\ldots,20^4$ we were able to estimate the critical coupling
for the percolation transition. This is worth studying, because
the percolation threshold is close to the chiral transition in
NC QED, and actually appears to coincide once dynamical
fermions are introduced. Although the relevance of monopoles to the
chiral transition is controversial \cite{Rak}, the percolation threshold is at
least a characteristic property of the ``QED vacuum'', and it will
prove interesting in this study.

Our results for the chiral transition are presented for each of the
four quenched models we have studied in the following table.
\begin{table}[h]
\begin{tabular}{rrrrr}
\hline
& \multicolumn{1}{c}{$\beta_c$} &
\multicolumn{1}{c}{$\delta$} & \multicolumn{1}{c}{$\beta_m$} &
\multicolumn{1}{c}{$\gamma$} \\
\hline\hline
NC QED & 0.257(1) & 2.1(1) & 0.86(3) & 1.03 \\
Block I & 0.664(4) & 2.3(1) & 0.82(6) & 1.07 \\
Block II & 0.112(4) & 2.3(1) & 0.96(7) & 1.25 \\
RL & 0.056(1) & 2.3(2) & 0.73(3) & 0.95 \\
\hline
\end{tabular}
\end{table}
The susceptibility exponent $\gamma$ was calculated using the scaling
relation $\gamma=\beta_m(\delta-1)$: it is included because the
quenched ladder approximation predicts it to be unity \cite{HKK}\cite{MBL}.
We see that the
main trend is for the models which are screened relative to the
standard formulation (ie. short wavelength modes less suppressed) to
break chiral symmetry at weaker bare coupling, whereas the less screened
model (Block II) has the transition at stronger coupling. The critical
inverse photon propagators $D_{\mu\nu}^{-1}(k)$ are shown for each
model in fig. 2, averaged over the corners of the Brillouin zone,
\begin{figure}
\vbox{
\psfig{figure=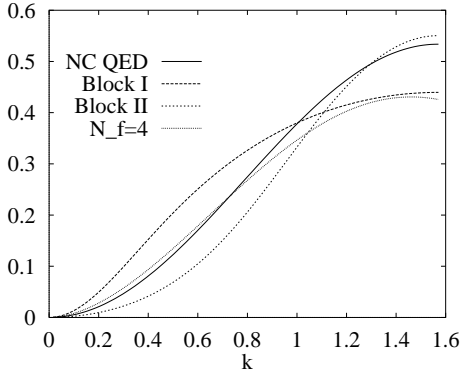,width=2.8in}
}
\caption{Critical inverse propagator $\beta_c D_{\mu\nu}^{-1}(k)$}
\end{figure}
since the blocking violates hypercubic symmetry to an extent. Also
shown for comparison is the critical inverse propagator for 4 flavor
QED, assuming a critical coupling $\beta_c\simeq0.205$ and logarithmic
screening using the perturbative formula. The crucial observation is
that all the curves differ strongly in the $k\to0$ limit, and approach
each other near the zone edge. This is conclusive evidence that chiral
symmetry breaking in QED is due to short-ranged interactions.
It is also striking how similar the estimates for the exponent $\delta$
are in the four models, even for the RL model which violates gauge
invariance. The universality of $\delta$ and (less convincingly)
$\beta_m$ remains to be understood -- perhaps it is a feature of the
staggered fermion formulation. It will be extremely valuable to get
quantitative results using other formulations.

In the next table we compare the critical couplings for the chiral and
monopole percolation thresholds.
\begin{table}[h]
\begin{tabular}{rrr}
\hline
& \multicolumn{1}{c}{$\beta_c(\chi{\rm SB})$} &
\multicolumn{1}{c}{$\beta_c$(percolation)} \\
\hline\hline
NC QED & 0.257(1) & 0.244(2) \\
Block I & 0.664(4) & 0.664(4) \\
Block II & 0.112(4) & 0.084(4) \\
RL & 0.056(1) & 0.060(1) \\
\hline
\end{tabular}
\end{table}
{}From this we learn that the two transitions can be made distinct if
screening is lessened: however for ``screened'' models (Block I,
$N_f=2,4$ QED) the two transitions appear to coincide \cite{Nf2}\cite{Nf4}.
Figure 2 also
supports a crude division of models into ``screened'' and
``unscreened''. In the quenched theory, we also see that
\begin{equation}
\beta_c(\chi{\rm SB})\geq\beta_c({\rm percolation})
\end{equation}
for {\sl all} models in which the photon fields are generated using a
gauge-invariant action.

The results of this study have raised more questions than have been
answered. Little support has been found for the scenario predicted by
the ladder approach to the gauged NJL model: we have been unable to
affect the critical exponents by tuning the short-ranged properties of
the model. Perhaps a more radical approach, such as a formulation in
momentum space \cite{Lag}, is needed.
We have also shown that monopole percolation
can be made separate from the chiral transition, at least in the
quenched case, but have found little evidence for the percolation
threshold occurring at a weaker coupling than the chiral transition,
except in the RL model. It would appear desirable to continue to
explore this problem from many different directions.

\end{document}